# Structural, magnetic and mechanical properties of 5 μm thick SmCo films for use in Micro-Electro-Mechanical-Systems

A. Walther[a,b], K. Khlopkov[c], O. Gutfleisch[c], D. Givord[a] and N.M. Dempsey[a]

[a]Institut Néel, CNRS-UJF, 25 rue des Martyrs, 38042 Grenoble, France

[b]CEA Léti - MINATEC, 17 rue des Martyrs, 38054 Grenoble, France

[c]IFW Dresden, Institute of Metallic Materials, Helmholtzstr. 20, 01069 Dresden, Germany

Abstract

5μm thick SmCo films were deposited onto Si substrates using triode sputtering. A study of the influence of deposition temperature ($T_{dep} \leq 600°C$) on the structural, magnetic and mechanical properties has shown that optimum properties (highest degree of in-plane texture, maximum in-plane coercivity and remanence (1.3 and 0.8 T, respectively), no film peel-off) are achieved for films deposited at the relatively low temperature of 350°C. This temperature is compatible with film integration into Micro-Electro-Mechanical-Systems (MEMS). The deposition rate was increased from 3.6 to 18 μm/h by increasing the surface area of the target from 7 to 81 cm$^2$. Mechanically stable films could be prepared by deposition onto pre-patterned films or deposition through holes in a mask.

# INTRODUCTION

Thick films (1-500 μm) of hard magnetic materials have many potential applications in magnetic MEMS [1]. Owing to their excellent intrinsic magnetic properties, rare-earth transition metal alloys ($Nd_2Fe_{14}B$ and SmCo based alloys) are very good candidate materials when high performance magnets are required [2]. The choice of material depends on the application requirements and processing constraints. From the point of view of the materials' intrinsic properties, NdFeB is the material of choice for high energy product films (theoretical $(BH)_{max}$ = 516 kJ/m$^3$; maximum value reported for a film is 400 kJ/m$^3$ [3]) while $SmCo_5$ is the material of choice for high coercivity films ($\mu_0 H_A$ = 40 T; maximum coercivity reported for a film is 5.6 T [4]). The choice of material may also be dictated by requirements for magnetic texture (i.e. a preference for the magnetic easy axes to be in-plane or out-of-plane). Under non-epitaxial conditions and for relatively thick films (≥ 1μm), NdFeB can be prepared with out-of-plane texture [3,5-7] while SmCo alloys can be prepared with in-plane texture [5,8,9]. Finally, the integration of such films into MEMS may dictate processing constraints such as a maximum processing temperature of 400°C for integration of films onto substrates which already contain integrated circuits ("above IC"). This limitation excludes NdFeB for this type of use, since it needs to be processed at higher temperatures (either deposited above 500°C [6] or annealed above 650°C [10]. However, integration of NdFeB is possible when the material is deposited/annealed before other temperature sensitive components. Note that the majority of published works on RE-TM films deal with films of thickness in the range of 100 nm - 1μm, which are too thin for use in MEMS. A recent review of hard magnetic materials for MEMS applications has been given in [11].

Three distinct hard magnetic phases of the SmCo system have been studied in film form: $SmCo_5$ [8,9,12,13], $SmCo_7$ [13,14] and $Sm_2Co_{17}$ [12]. Enrichment in Co with respect to the $SmCo_5$ phase leads to an increase in magnetisation and thus energy product but a loss in magnetocrystalline anisotropy [2]. Thus for applications in which very large coercivites are not needed, $SmCo_7$ and $Sm_2Co_{17}$ are the phases of choice. In-plane textured $SmCo_7$ or $Sm_2Co_{17}$ films have been used for biasing of yttrium iron garnet (YIG) [15], permalloy [16] and Colossal Magneto Resistance (CMR) [17] films. More recently, two prototype micro-actuators have been made with patterned isotropic $Sm_2Co_{17}$ films [18]. To our knowledge, the mechanical properties of RE-TM films have not been reported to date. In this paper, a study of the structural, magnetic and mechanical properties of $SmCo_7$ films destined for use in MEMS will be presented. Si substrates have been used for compatibility with standard micro-technology processing, in contrast with the above cited studies on the applications of such films [15-18], in which oxide substrates (typically $Al_2O_3$) were used.

**EXPERIMENTAL DETAILS**

A $Sm_{23}Co_{77}$ cast target of diameter 3 cm, was sputtered using a triode sputtering system (Ar pressure $\approx 2\times10^{-3}$ mbar). Cr(50 nm) / SmCo (5 µm) / Cr (50 nm) films were deposited at a rate of 3.6 µm/hour onto Si substrates at temperatures $\leq 600°C$ (note that even when no power is sent to the substrate furnace, the substrate temperature gradually rises during deposition, reaching a final value of about 200°C – films deposited under such conditions are labeled "cold" deposited). Post-deposition annealing was carried out ex-situ in a vacuum of $10^{-5}$ mbar. XRD structural characterization was carried out in both θ-2θ and pole figure configurations using Co and Cu radiations, respectively. Observations and analysis were made with an FEG-SEM (Gemini 1530) equipped with an EDX detector. Room temperature magnetic properties were measured using a vibrating sample magnetometer. Film curvature was measured using a profilometer (Dektak).

**RESULTS AND DISCUSSION**

*Structural properties*

The θ-2θ XRD patterns of the as-deposited films are shown in figure 1. For $T_{dep} \leq 300°C$, the films appear to be amorphous (only Si and Cr peaks observed). Additional diffraction peaks of the films deposited at 350°C and 400°C may be indexed as the $TbCu_7$ structure while the films deposited at 500°C have peaks of both the $TbCu_7$ and $Th_2Zn_{17}$ structures. Munakata et al. also observed the transition to a Sm poorer phase on increasing the deposition temperature, though in their case the shift was from $SmCo_5$ to $Sm_2Co_{17}$ [19]. The position of the (hk0) peaks of the $TbCu_7$ phase, which are dominant because of in-plane texture (see below), was found to vary with deposition temperature, initially shifting to higher angles when the deposition temperature increases from 350 to 400°C and then shifting to lower angles when the deposition temperature is increased to 500°C. The value of the "a" lattice parameter was estimated to be 4.9Å, 4.82Å and 4.9Å for $T_{dep}$ = 350°C, 400°C and 500°C, respectively, compared to a=4.997 Å for the $SmCo_5$ phase [20]. This variation with deposition temperature is attributed to an initial reduction in Sm content of the $TbCu_7$ phase (presumably due to Sm evaporation at the higher deposition temperature) and then a re-enrichment of the Sm content of the $TbCu_7$ phase caused by the appearance of the $Th_2Zn_{17}$ phase which is richer in Co. Note that while EDX analysis indicates an average film composition of $Sm_{14}Co_{86}$, the resolution of our EDX analysis was not precise enough to verify the assumption concerning the overall shift in composition as a function of the deposition temperature.

Comparison of the relative intensities of the diffraction peaks of the θ–2θ scans indicates that the c axes of the $TbCu_7$ phase are preferentially oriented in the plane, with the degree of texture decreasing with increasing deposition temperature. The in-plane texture was further evidenced using pole-figure analysis (figure 2). While the films deposited at 350°C have strong in-plane texture, as revealed by the sharpness of the (200) pole figure, the films deposited at 500°C are only weakly textured, with the (1 1 0) pole figure being sharper than the (200) pole figure. In contrast with the strong (200) texture observed here, several literature reports refer to (110) texture for $SmCo_5$ [19,21] and $SmCo_7$ [14,22]. Though there have been studies reporting the influence of single parameters on the predominance of (200) or (110) texture (e.g. oxygen level [23], sputtering pressure [13,22], deposition rate [24,25]), the texture is most probably determined by a combination of many factors. Indeed the cited reports on the influence of deposition rate are contradictory [24,25]. Broadening of the XRD peaks was observed with increasing deposition temperature and may be due to a combination of grain size reduction (it is plausible that the grain size of the 1:7 phase may decrease when grains of the 2:17 phase form) and strain. Considering only grain size, Scherrer formula analysis of Gaussian fits of the (2 0 0) peak of the 1-7 phase indicates that the crystallite size of this phase decreases with increasing deposition temperature, from 16.5 nm at 350°C to 5 nm at 500°C. However the validity of these estimates is questionable because of the high strain levels measured in these films (see below).

AFM analysis indicates that the films deposited up to 400°C are very smooth (RMS roughness ≈ 2 nm) while those deposited at 500°C are rougher with an RMS ≈ 10 nm (images not shown). In most cases, attempts to observe the grain structure by high resolution SEM-FEG imaging of fracture surfaces of the films' cross sections were unsuccessful, presumably owing to the absence of brittle fracture. Nevertheless, some microstructural features were observed in the film deposited at 500°C. Roughly equiaxed grains with diameters of the order of 80nm were observed on an internal surface which was revealed by film fracture (presumably the void was formed due to the presence of a dust particle on the substrate) (Figure 3a). A grain structure with comparable feature size was observed in plane-view images of the film after an Ar ion etch of the Cr capping layer (Figure 3b,c). The difficulty in observing the microstructure of our SmCo$_7$ films is coherent with the fact that clearly resolved images of typically oblong grains have only been reported for SmCo$_5$ films [13,19,26].

*Magnetic properties*

The magnetic properties were found to depend strongly on the deposition temperature. Films deposited at temperatures ≤ 300°C (data not shown) are magnetically soft, in agreement with the absence of x-ray diffraction peaks in these films. The in-plane (ip) and out-of-plane (oop) hysteresis loops of films deposited in the range 350-500°C are shown in figure 4a-c. The degree of magnetic texture, as characterized by the difference between the ip and oop hysteresis loops, was found to decrease with increasing deposition temperature. In agreement with the XRD data, in-plane texture is found (i.e. the magnetically easy c-axes are preferentially aligned in the plane of the film). For the best textured film ($T_{dep}$=350°C), $M_r(ip)/M_{8T}(ip) \approx 0.83$, $M_r(ip)/M_r(oop) \approx 12$ and the absolute value of remanence is estimated to be 0.8T. For a uniform distribution of the c-axes within the film plane, $M_r(ip)/M_s(ip) = 2/\pi \approx 0.64$ for a system of non-coupled grains. Thus, the high value of $M_r(ip)/M_{8T}(ip)$ measured in this system, indicates that the grains are exchange coupled. The shift of the out-of-plane loop indicates that not all grains have been reversed under the maximum applied field of 8T. The in-plane loop of the sample deposited at 500°C is characterized by the presence of two phases (distinguished because of their different values of coercivity), in agreement with the XRD data (i.e. SmCo$_7$ and Sm$_2$Co$_{17}$). The coercivity of the films in their as-deposited state drops from a value of 1.3T for $T_{dep}$ = 350°C to a value of 0.5 T for $T_{dep}$ = 500°C. Thus, 350°C appears to be the optimum deposition temperature as it gives the maximum values or coercivity and remanence and the films produced at this temperature have an estimated energy product of 112 kJ/m$^3$.

Post-deposition annealing at a temperature of 750°C for 10 minutes leads to crystallization and thus the development of coercivity in samples deposited at $T_{dep} \leq 300°C$ and to an increase in coercivity and squareness for films which were deposited in the crystallized state (Figure 5). The improvement in squareness leads in an increase in the energy product of the film deposited at 350°C from 112 J/m$^3$ in the as-deposited state to 144 kJ/m$^3$ in the annealed state. This energy product is to be compared with values of 90 kJ/m$^3$ reported for isotropic $Sm_2Co_{17}$ films [18], 120 kJ/m$^3$ reported for textured $SmCo_5$ [13] and $SmCo_7$ [14] films and 206 kJ/m$^3$ reported for textured $Sm_2Co_{17}$ films [27], all of which have thicknesses comparable to the films studied here.

*Influence of target size*

A benefit of working with the triode sputtering technique is that the target size can be easily varied. Following the above study, a larger target (9 x 9 cm$^2$) of the same composition was used so as to increase the deposition rate and to extend the zone of homogeneous film thickness on the substrate. This increase in target surface area (from 7 cm$^2$ to 81 cm$^2$) raised the deposition rate from 3.6 µm/h to 18 µm/h. This deposition rate is significantly higher than the maximum values previously reported for SmCo films (5 µm/h [9,13], 10 µm/h [18]). Furthermore, the high rates of deposition are achieved over relatively large areas: films deposited on 100 mm Si wafers are only 20% thinner at the perimeter than at the centre. The deposition rate and zone of homogeneous thickness are important factors for assessing the suitability of a given technique for integrating films into MEMS.

Film deposition with the large target at the temperature which proved to be optimum for the small target (350°C) produced samples with relatively poorer properties (lower coercivity and lower squareness) and it was found that the magnetic properties could be improved by increasing the deposition temperature to 400°C (figure 6). An in-plane energy product of 140

kJ/m$^3$ was reached for the optimum deposition temperature (400°C). The degree of (200) texture was found to decrease with increasing deposition temperature, in agreement with the results obtained with the small target (Figure 7). A comparison of the results for the two different target sizes reveals an influence of deposition rate on growth mechanisms.

*Mechanical properties*

The mechanical properties of the as-deposited films were found to depend on the substrate temperature during film deposition. While films deposited at temperatures of up to 400°C remained entirely adhered to the Si substrate, small pieces (typically a few mm$^2$) of those deposited at 500°C peeled off and finally films deposited at 600°C almost entirely peeled off. All as-deposited films were found to be under tensile stress and the resulting curvature was measured using a profilometer on samples of typical size 1 cm$^2$. The values of curvature were then used to estimate the residual tensile stress using the Stoney formula [28]. Tensile stress as a function of deposition temperature for the films deposited with the small target is plotted in figure 8. The size of the error bars indicates the difference in stress values measured along orthogonal directions.

Post-deposition annealing led to film peeling for all deposition temperatures. This peel-off is attributed to the difference in thermal expansion coefficients of the SmCo layer and the Si wafer and is the reason why most studies of thick films were made on substrates other than Si [9, 15, 18] (thermal expansion coefficients of 6x10$^{-6}$/K and 12 x10$^{-6}$/K have been measured parallel and perpendicular to the c-axis of SmCo$_5$ at 600°C [29] while Si has a value of 2.6 x10$^{-6}$/K). Peeling was found to be related to the lateral dimensions of the film. When 5 µm thick SmCo films were deposited onto pre-patterned Si/SiO wafers which have topographic relief (trench motifs with individual trenches of depth 5 µm with trench/wall widths on the scale of 5-100 µm), peel off did not occur from film sections with at least one dimension on the scale 5-100 μm. Furthermore, films of thickness up to 20 μm deposited onto Si through a mask with 4 x 4 mm sized holes remained adhered to the substrate and were crack free. These results are reminiscent of those obtained on NdFeB films [3].

**CONCLUSIONS**

Triode sputtering has been used to deposit thick SmCo films (5 – 20 μm) onto Si substrates. Deposition rates of 3.6 and 18 μm/h have been achieved for target surface areas of 7 cm$^2$ and 81 cm$^2$, respectively. A study of the influence of deposition temperature reveals that relatively low deposition temperatures (350-400°C) are optimal with respect to structural, magnetic and mechanical properties (highest degree of in-plane texture, maximum coercivity and maximum remanence, no film peel-off). These relatively low temperatures are compatible with "above IC" processing, and establish the potential for the integration of such films into MEMS. Film peel-off due to strain can be avoided by the preparation of films with reduced lateral dimensions (i.e. deposition through masks or onto pre-patterned substrates).

**ACKNOWLEDGMENTS**

This work was carried out in the framework of the ANR "Nanomag2" project. The authors would like to thank A. Lienard and L. Ortega for technical assistance. K. K. is grateful for financial support from DFG (SFB 463) and O. G. for a CNRS visiting research fellowship at Institut Néel.

**FIGURE CAPTIONS**

Figure 1: θ-2θ XRD patterns of the as-deposited SmCo films as a function of deposition temperature.

Figure 2: (110) and (200) pole figures of SmCo films deposited at (a) 350°C and (b) 500°C. Note that the maximum intensity of the (2 0 0) pole figure of the sample deposited at 350°C is 10 times greater than that of the other pole figures.

Figure 3: SEM-FEG images of a SmCo film deposited at 500°C taken with an in-lens detector: (a) cross sectional image of a fractured film; (b,c) plane-view images after Ar ion etching of the Cr capping layer.

Fig 4: In-plane (ip) and out-of-plane (oop) hysteresis loops of SmCo films deposited in the temperature range 350-500°C.

Figure 5: In-plane hysteresis loops of films annealed at 750°C for 10 minutes as a function the temperature at which the films were deposited.

Figure 6: Comparison of the in-plane hysteresis loops of SmCo films deposited in the temperature range 350 - 400°C using a large (9x9 cm$^2$) target.

Figure 7: Comparison of the $\phi-2\phi$ xrd patterns of SmCo films deposited in the temperature range 350 - 400°C using a large (9x9 cm$^2$) target.

Figure 8: Influence of deposition temperature on the tensile stress of SmCo films. The size of the error bars indicates the difference in stress values measured along orthogonal directions.

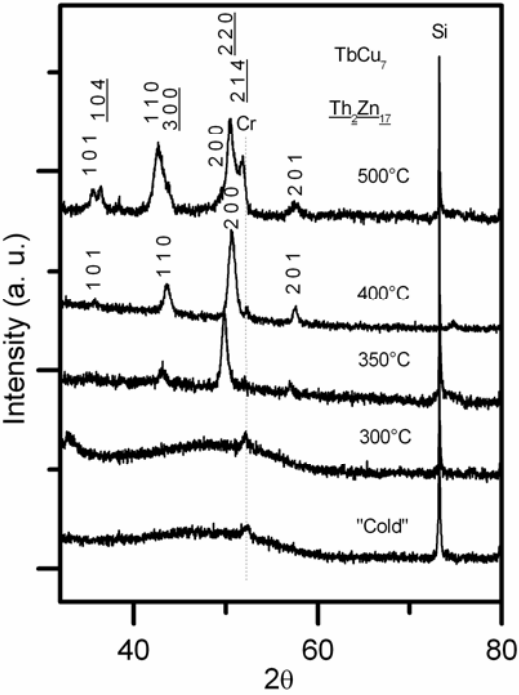

Figure 1

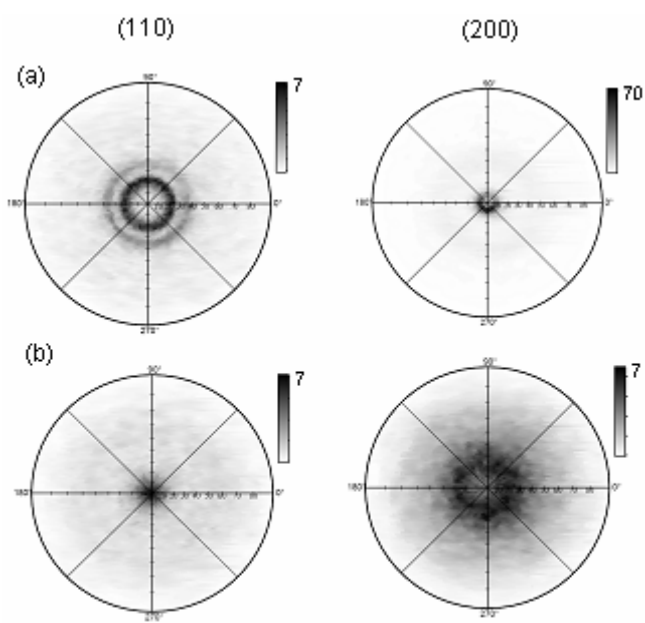

Figure 2

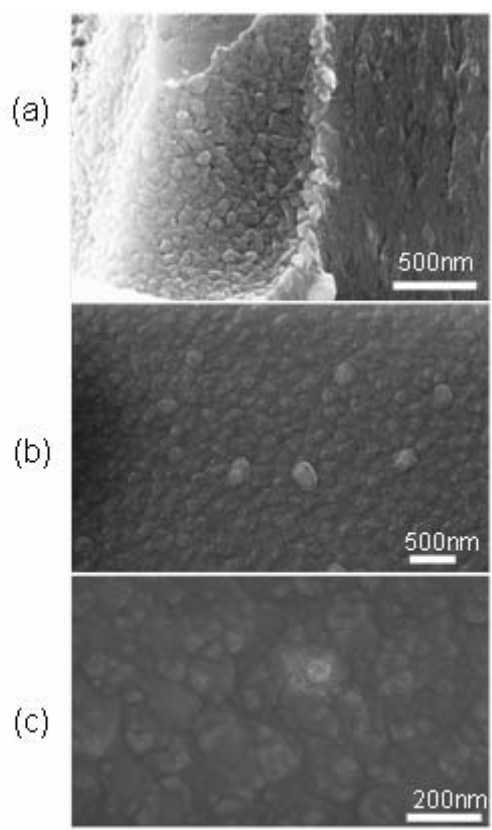

Figure 3

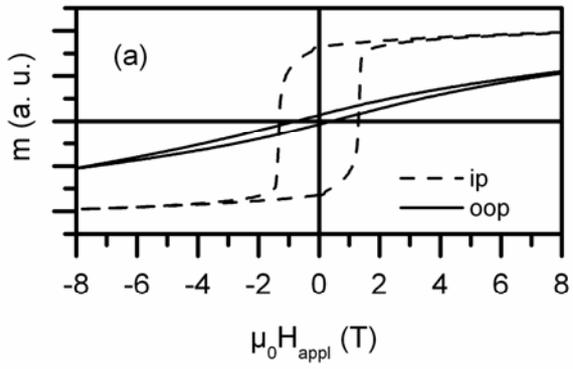

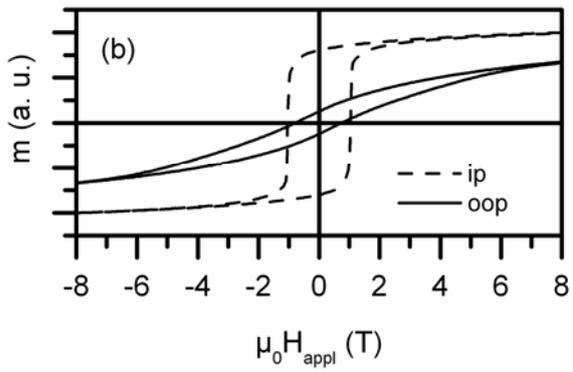

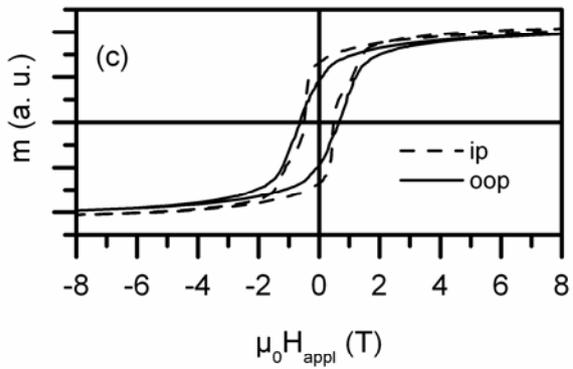

Figure 4

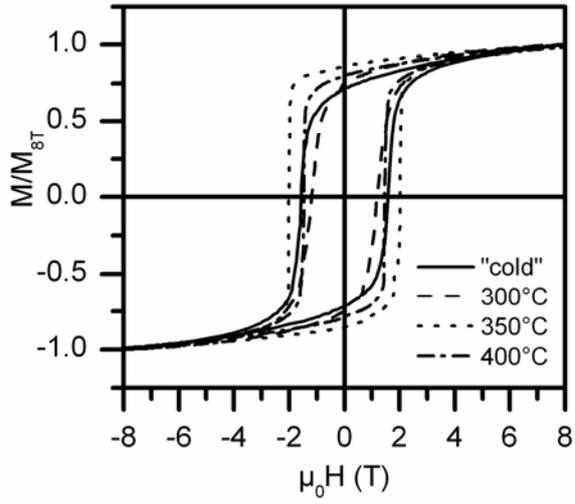

Figure 5

Figure 6

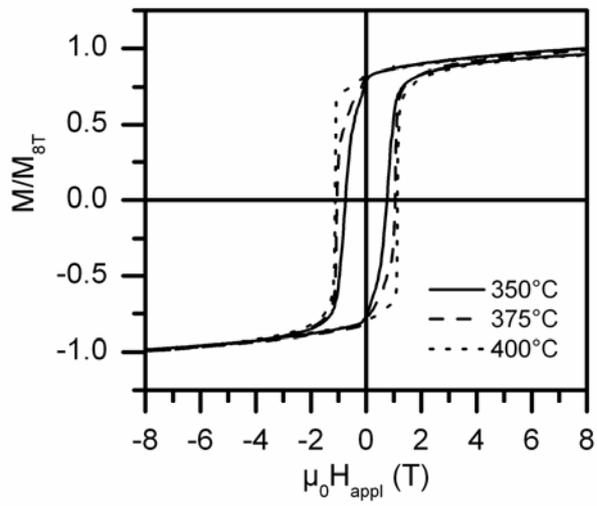

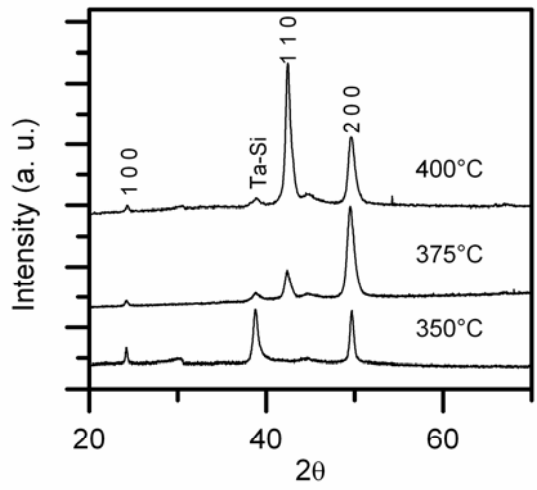

Figure 7

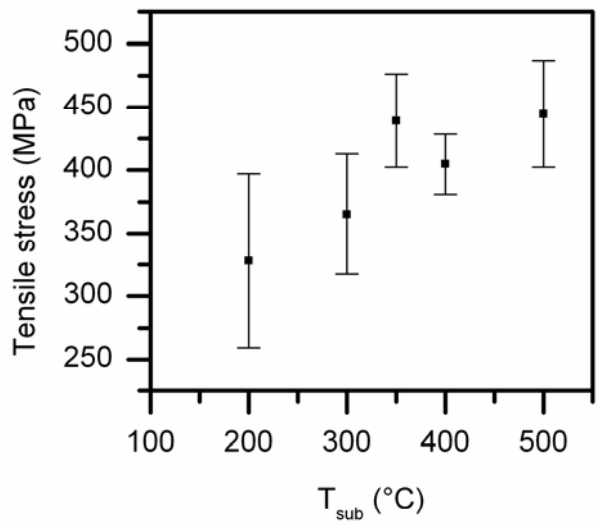

Figure 8